\documentclass{article}

\usepackage[final]{nips_2018}

\usepackage[utf8]{inputenc} 
\usepackage[T1]{fontenc}    
\usepackage{hyperref}       
\usepackage{url}            
\usepackage{booktabs}       
\usepackage{amsfonts}       
\usepackage{nicefrac}       
\usepackage{microtype}      
\usepackage{comment}
\usepackage{graphicx}
\usepackage{caption}
\usepackage{subcaption}

\PassOptionsToPackage{number}{natbib}
\setcitestyle{notesep={; },round,aysep={},yysep={;}}

\title{Looking Deeper into Deep Learning Model: Attribution-based Explanations of TextCNN}

\newcommand\Mark[1]{\textsuperscript#1}

\author{
  Wenting Xiong\Mark{1$^*$}, Iftitahu Ni'mah\Mark{1$^{\dag}$}\thanks{equal contribution}, Juan M. G. Huesca\Mark{1}\\
  \textbf{Werner van Ipenburg\Mark{2$^{\ddag}$}, Jan Veldsink\Mark{2$^{\ddag}$}, Mykola Pechenizkiy\Mark{1$^{\dag}$}}\\
  \Mark{1}Eindhoven University of Technology, the Netherlands \\
  \Mark{2}Cooperatieve Rabobank U.A.\\
  \Mark{\dag}\texttt{\{i.nimah, m.pechenizkiy\}@tue.nl} \\
  \Mark{\ddag}\texttt{\{werner.van.ipenburg, jan.veldsink\}@rabobank.nl}
}

\begin{document}

\maketitle

\begin{abstract}
 Layer-wise Relevance Propagation (LRP) and saliency maps have been recently used to explain the predictions of Deep Learning models, specifically in the domain of text classification. Given different attribution-based explanations to highlight relevant words for a predicted class label, experiments based on word deleting perturbation is a common evaluation method. This word removal approach, however, disregards any linguistic dependencies that may exist between words or phrases in a sentence, which could semantically guide a classifier to a particular prediction. In this paper, we present a feature-based evaluation framework for comparing the two attribution methods on customer reviews (public data sets) and Customer Due Diligence (CDD) extracted reports (corporate data set). Instead of removing words based on the relevance score, we investigate perturbations based on embedded features removal from intermediate layers of Convolutional Neural Networks. Our experimental study is carried out on embedded-word, embedded-document, and embedded-ngrams explanations. Using the proposed framework, we provide a visualization tool to assist analysts in reasoning toward the model's final prediction.

\end{abstract}

\section{Introduction}

Convolutional Neural Networks (CNNs) have been showing promising results in text classification, including movie reviews binary classification, multi-class classification of the sentiment treebank, and topic categorization ~\citep{Collobert2011Natural, kim2014convolutional, conneau2017very}. This competitive performance of CNN on a wide range of text classification tasks has become its main attraction as end-to-end applications in industries beyond computer vision applications. However, in many critical domains (e.g.\ banking, health care and medical services), there is also an increasing demand for models and an evaluation framework that can support aspects of CNN models interpretability and exploratory analysis. 

The importance of model interpretability in the domain of banking services is exemplified in the deployment of machine learning models for analyzing customer behaviour in the Customer Due Diligence (CDD) stage of Know Your Customer (KYC). Given customer data in a form of CDD reports and the corresponding historical assessment from the analyst (labels of customer categorization), a classifier can be built to characterize customers based on the content of their reports, e.g.\ as a category of ``low'' or ``high'' financial risk customer. Providing an interpretable model is therefore desirable since it could reveal any confounding factors that further explain the model's final prediction. For instance, by providing the reasoning why a customer is categorized as ``high'' risk, instead of ``low'' one -- or the reasoning why the model misclassifies a customer during the validation stage. 


Several approaches have been explored for improving interpretability of Deep Neural Network (DNN) models. Proposed approaches so far include global (layer-wise) and local (individual feature importance) explanation methods, as exemplified in the preliminary work on visualizing DNN for image classification \citep{simonyan2013deep, samek2017evaluating, ancona2018towards}. The latter work summarizes several attribution methods for explaining what DNN models have learned in the corresponding prediction task, including the two back-propagation-based methods, i.e. Layer-wise Relevance Propagation (LRP) and saliency maps. An evaluation metric based on the sensitivity analysis for evaluating different gradient-based and perturbation-based methods for image and text classification was proposed in \citep{ancona2018towards}. 

In the domain of text classification, the aforementioned attribution methods were also employed to further explain the predictions of neural models. The works on local explanation \citep{nguyen2018comparing} and visualization of linguistic compositionality in neural models \citep{li2016visualizing} utilized the first derivative saliency to identify most influential inputs (words) for and against a particular prediction. Likewise, LRP was also employed for explaining CNN predictions on a topic categorization task \citep{arras2016explaining}.

To compare different attribution-based models, experiments with word removal were used in ~\citep{arras2017relevant}. The main idea is that by deleting the words with the highest attribution scores, a drastic drop in the model accuracy should be observed. However, there is also a drawback. There may exist dependent factors that contribute to the change of accuracy scores. For instance, the model's decisions could be influenced by the relevance of phrases ($n$-grams). Removing words will not only eliminate the contribution of the particular words, but could also affect the contribution of other words within the same context window ($n-$grams), sentence, or document. 

In this paper, we employ the two attribution methods (i.e.\ saliency maps and LRP) on binary and multi-class classification of customer reviews (public data sets). Different from previous approaches that measure the quality of explanation methods with ``word deleting'' perturbation experiments, we evaluate the attribution scores with ``feature removal'' method. As example of an application in real world data set, we utilize our CNN model and the two attribution-based explanations on CDD reports (corporate data set). We also developed an interactive visualization tool \footnote{https://peaceful-journey-19056.herokuapp.com/} to further help analysts in investigating the model's prediction outputs.

The rest of paper is organized as follows. In Section~\ref{sec:bg-cnn}, we describe the architecture of CNN in this study. The two attribution-based explanations and our proposed evaluation framework are explained in Section~\ref{sec:bg-attribution}. Experiments and results are discussed in Section~\ref{sec:experiments}. The conclusion is presented in Section ~\ref{sec:conclude}.

\section{CNN model}
\label{sec:bg-cnn}

\begin{figure}[!ht]
\centering
\includegraphics[scale=0.6]{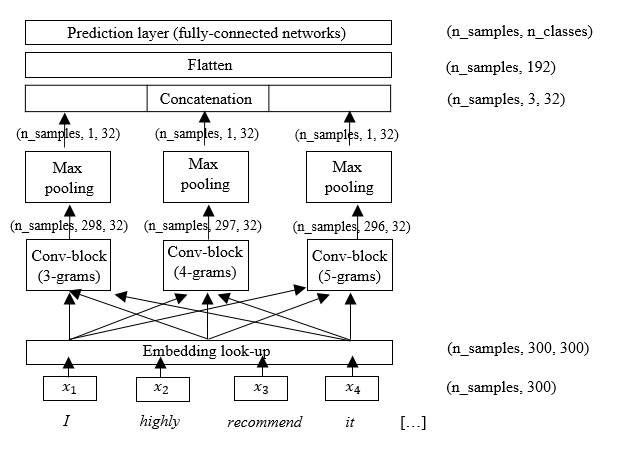}
\caption{Architecture of the TextCNN}
\label{fig:CNNmodel}
\end{figure}

We employed a word-based CNN model as a document classifier, i.e.\ to predict whether the text review is positive or negative (binary classification task) and to perform the categorization of text documents (multi-class classification). Figure~\ref{fig:CNNmodel} depicts CNN architecture in this study, which we refer as TextCNN. ``Conv-block'' denotes the convolutional layer with the corresponding feature map (filter). In image classification problem, the filters correspond to red, green, blue (RGB) filters, while in this text classifier the filters are referred to three (3) different $n-$grams filters, (where $n=3,4,5$ in this study).

\section{Attribution-based explanations}
\label{sec:bg-attribution}

\paragraph{Saliency maps}
Gradient-based saliency maps or Sensitivity Analysis (SA) \citep{simonyan2013deep} construct the attribution score by taking the partial derivative of the target output for a particular class $c$ ($f^c$) with respect to the input features $x$. Instead of the common absolute form of saliency, we employed raw values of saliency (signed saliency), as in: \[R_j^c = \frac{\partial }{{\partial {x_{j}}}}{f^c}\left( x \right)\]
\paragraph{LRP}
LRP~\citep{samek2017evaluating} redistributes the prediction score $f_c(x)$ layer by layer until reaching the desired layer:
\[R_j^c = \sum\limits_k {\frac{{{x_j}{w_{jk}}}}{{\sum\limits_j {{x_j}{w_{jk}}} }}R_k^c} \]
The following rule holds for LRP attribution scores from all layers that for a particular class $c$, the sum of attribution scores on a layer is equal to the prediction score $f_{c}(x)$:\[\sum\limits_i {R_i^c}  = ... = \sum\limits_j {R_j^c}  = \sum\limits_k {R_k^c}  = ... = {f_c}\left( x_{ij} \right)\]
\section{Evaluation framework}

\subsection{Embedded-word relevance}
\label{eval1}

In this experiment, the attribution score that is assigned on each feature (each dimension) of word embedding was utilized without any perturbation-based experiments. For both the quantitative and qualitative evaluation of the embedded-word relevance, we carry out experiments with document highlighting ~\citep{arras2016explaining}, i.e. by using the document embedding as the input of classifiers (KNN, SVM, Decision Tree, Random Forest) to predict the category label of the corresponding embedded document. A higher accuracy is expected for weighted document representations if truly important features are assigned higher weight. We do not employ ``feature deletions'' in this word-based relevance model since we are more interested in higher abstraction than words (i.e. perturbations of features of embedded-$n-$grams or embedded-document) as explained in section~\ref{eval2} and ~\ref{eval3}.

Given a three-dimensional output of the embedding layer ($n$-samples ($i$), $n$-sequence of words ($j$), dimension of word embedding ($k$)), the attribution score is assigned for each $k$ dimension of this matrix. To create a document representation (document embedding), the attribution score of each word is used as weighting factor. The feature-based attribution score for word-$j^{th}$ in document-$i$ and embedding column $k$ is described as $R^{c}_{ijk}$, while the total attribution score for this word-$j^{th}$ is $\sum_k R^{c}_{ijk}$. Given the representation of words (word embedding) $e(w_j) = v^{(0)}, v^{(1)}, \cdots, v^{(k)}$, the non-weighted document representation for document-$i$ is the average of representation of words in that document $\frac{1}{j}\sum_{j}e(w_j)$. The weighted document representation for document-$i$ is $\frac{1}{j}\sum_{j}\left ( \sum _{k}R_{ijk}^{c} \right )e(w_j)$.

\subsection{Embedded-document relevance}
\label{eval2}
Experiments based on the embedded-document perturbations were performed to evaluate whether the important features are assigned high attribution scores. Intuitively, different fragments of a document (e.g.\ between sentences) may tell different sentiment polarity weights. A review could be started by mentioning a negative criticism about a small aspect of a product, but the final conclusion may give positive recommendation. Assuming these different aspects of polarities are embedded as features of the learned document embedding, we utilize ``feature'' or each dimension of the embedded document to evaluate the importance of scores assigned by attribution methods in the corresponding prediction task. 

Similar to the score acquired in Section~\ref{eval1}, the feature-based attribution score for word-$j^{th}$ in document-$i$ and embedding column $k$ is described as $R^{c}_{ijk}$, while the total attribution score for this word-$j^{th}$ is $\sum_k R^{c}_{ijk}$. The attribution score for each embedding column of document embedding $e(x_i)$ is calculated by adding the relevance score of words in that document $\sum_{j} R^{c}_{ijk}$. The feature removal was done by setting all values in the corresponding columns to be 0. The evaluation was carried out on three (3) different settings:

\begin{enumerate}
    \item \textit{Removing features with the largest attribution scores.}
The embedding columns with the largest attribution scores for the true class were removed. The accuracy was therefore expected to be lower. For a correctly classified document, the predicted probability for its true class should be lower. 
    \item \textit{Removing features with the smallest attribution scores.}
The embedding columns with the smallest absolute attribution scores for the true class were removed. For both methods, the predicted probability should not be affected more than by randomly removing an embedding column. The purpose of this evaluation was to assess whether features with low attribution scores are truly unimportant features. 
    \item \textit{Removing features that contribute differently for different classes.}
For a document $x_i$, the attribution difference between true class $c$ and class $c'$ for embedding column $k$ is $\sum_{j} (R^{c}_{ijk} - R^{c'}_{ijk})$. When the columns with the largest attribution differences were removed, the predicted probability for class $c$ should decrease while the probability for class $c'$ should increase. This setting was only applied to classification tasks with multiple classes. 
\end{enumerate}

\subsection{Embedded-ngrams relevance}\label{eval3}

In our TextCNN model, the learned feature representation from convolutional layer hypothetically represents the $n$-gram features. For each filter, only the convolution window with the maximum value has an impact on the output (after a max pooling layer). Thus, we assume that removing a filter on a convolutional layer is equivalent to removing representation of an $n$-gram feature. Here, we defined a filter of a convolutional layer as a ``feature''.  Each filter was assigned by one non-zero attribution score, which represents attribution score of the $n-$grams of the input sequence. Likewise, the evaluation was conducted on three different settings as previously explained in section~\ref{eval2}. 

\section{Experiments and analysis}
\label{sec:experiments}

\subsection{Data sets}

Table~\ref{tab:datasets} shows three data sets that were used in this study and their corresponding statistics. TextCNN was trained on these three datasets. The corresponding classification performance is shown in Table~\ref{training_acc}. 

\paragraph{Yelp reviews (public data set)} The data set \footnote{https://www.yelp.com/dataset} is a collection of customer reviews on Yelp. For every review text, the customer gave it a ``stars`` rating ranging from $1$ to $5$. A higher rating indicates a more positive review. On this Yelp review data set, we removed neutral reviews with 3 stars. We redefined the reviews labeled as $1$ and $2$ to label $0$, and the reviews labeled as $4$ and $5$ as label $1$. As a result, the classification task on Yelp review data set was binary. 
\vspace{-1em}
\paragraph{US consumer finance complaints (public data set)} The dataset \footnote{https://www.kaggle.com/cfpb/us-consumer-finance-complaints} contains the customer complaints about 11 financial products and services. Each complaint contains one or more sentences.
\vspace{-1em}
\paragraph{Customer Due Diligence (CDD) reports (corporate data set)} is an extracted report of customers from Customer Due Diligence (CDD) cases. This data set contains pre-processed text reports with the corresponding risk-based labels, i.e. whether the customer is categorized as ``low'' (class ``0'') or ``high'' (class ``1'') financial risk.

\begin{table}[!ht]
\vspace{-2em}
  \caption{Datasets used in this study}
  \label{tab:datasets}
  \centering
  \begin{tabular}{lcccc}
    \toprule
         & Corpus & Average Length  & Shortest & Longest \\
         & size & (nr. of words)  & Length & Length \\
    \midrule
    Yelp reviews & 55.790  & 22 & 6 & 22    \\
    Consumer complaints & 64.821 & 198 & 13 & 912 \\
    CDD reports & 961 & 2.635  & 862  & 6.219  \\
    
    \bottomrule
  \end{tabular}
\end{table}

\begin{table}[!ht]
\vspace{-1em}
\centering
\caption{Performance of the trained TextCNN model}
\begin{tabular}{@{}lccccc@{}}
\toprule
Dataset & epochs & loss   & accuracy (\%) & \begin{tabular}[c]{@{}c@{}}validation\\loss\end{tabular} & \begin{tabular}[c]{@{}c@{}}validation\\ accuracy (\%)\end{tabular} \\ \midrule
Yelp reviews & 3 & 0.0788 & 97.6   & 0.1375  & 95.27 \\
Consumer complaints & 15 & 0.3314 & 89.5   & 0.6031  & 85.45 \\
CDD reports & 10 & 0.0867 & 98.57 & 0.1492 & 94.82 \\
\bottomrule
\end{tabular}
\label{training_acc}
\end{table}

\subsection{Evaluating embedded-word relevance}

\begin{figure}[!ht]
\centering
\begin{subfigure}[t]{\textwidth}
\includegraphics[scale=.65]{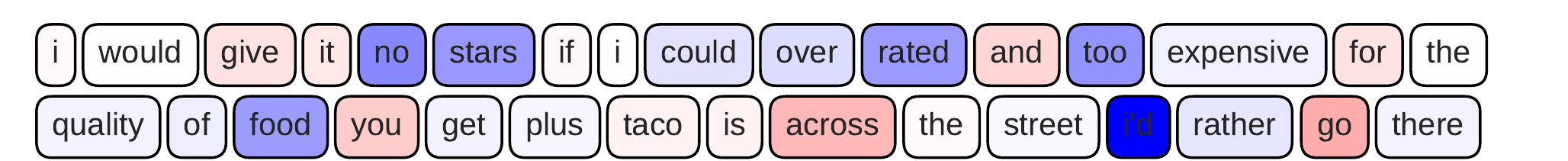}
\caption{LRP}
\end{subfigure}
\begin{subfigure}[t]{\textwidth}
\includegraphics[scale=.65]{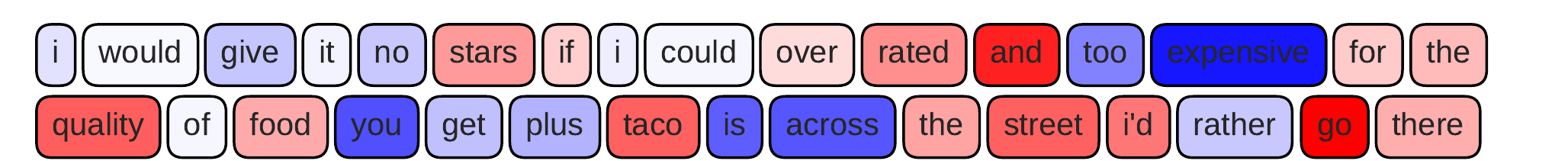}
\caption{Saliency maps (SA)}
\end{subfigure}
\caption{Visualization of attribution scores on a negative review (correctly classified class ``0'')}
\label{fig:att_vis0}
\end{figure}

\begin{figure}[!ht]
\centering
\begin{subfigure}[t]{\textwidth}
\includegraphics[scale=.65]{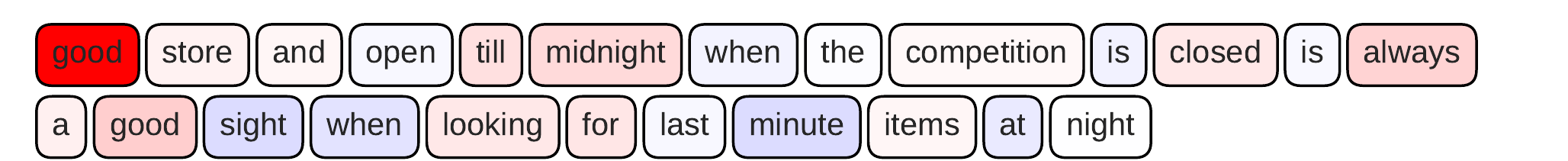}
\caption{LRP}
\end{subfigure}
\begin{subfigure}[t]{\textwidth}
\includegraphics[scale=.65]{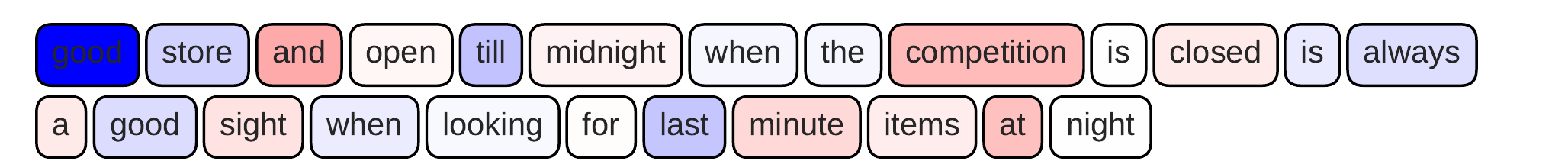}
\caption{Saliency maps (SA)}
\end{subfigure}
\caption{Visualization of attribution scores on a positive review (correctly classified class ``1'')}
\label{fig:att_vis1}
\end{figure}

Figures~\ref{fig:att_vis0} and ~\ref{fig:att_vis1} show the visualization of the two attribution methods on the correctly classified ``0'' and ``1'' of Yelp reviews respectively. Positive scores (positive contribution to class ``1'') are shaded in ``red'', while negative scores (negative contribution to class ``1'') are highlighted as ``blue''. From Figure~\ref{fig:att_vis0}, we can see that LRP was able to highlight the compositionality of negative words (e.g.\ ``no stars'') that contributes to negative ``0'' class. SA could find a negation (``no'' word), but not as a phrase or combined words. Both attribution methods were able to put relevance scores on phrase with excessive expression (``too"), but SA put a higher weight on this type of phrase. In the example of positive review (Figure~\ref{fig:att_vis1}), LRP assigned a higher relevance score on positive words (e.g. ``good''), while in this example, SA did not correctly assign the score on the same word or phrase as compared to LRP.      

For measuring the quality of the embedded-word relevance scores, we employed different weighting schemes of document embedding (i.e.\ based on the score assigned after embedding layer) as an input of a classifier. The comparison on four classifiers is shown in Table~\ref{acc_rep}. ``w-0'' denotes unweighted document embedding as input, ``w-LRP'' denotes LRP-based weighted, and ``w-SA'' denotes saliency-based weighted document representation. On three data sets, the LRP-based weighted document representation achieved higher accuracy as compared to the non-weighted and saliency-based weighted ones. In this experiment, with the LRP attribution as weighting factor, words that are relevant to the actual class label were assigned larger weights, and thus became more influential in the generated document representations. While saliency-based weighting (w-SA) is not always distinctive, as such, the classification performance is often similar or even lower than non-weighted document embedding.

\begin{table}[!ht]
\centering
\caption{Accuracy score (\%) (using document embedding as input of classifier)}
\resizebox{\textwidth}{!}{
\begin{tabular}{@{}l ccc ccc ccc@{}}
\toprule
Classifier & \multicolumn{3}{c}{Yelp reviews} & \multicolumn{3}{c}{US Customer complaints}  & \multicolumn{3}{c}{CDD reports} \\ 
 & w-0 & w-LRP  & w-SA & w-0 & w-LRP  & w-SA & w-0 & w-LRP  & w-SA \\ 
\midrule
KNN & 72.5 & \bf 92.75  & 69.25 & 25.74 & \bf 58.38 &  58.38 & 94.59 & \bf  100 & 94.59 \\
SVM &  51.25 & \bf 91.5 & 65.25 & 12.87 & \bf 23.65 &  23.65 & 54.05 & \bf  100 & 54.05 \\
Decision tree & 69.5 & \bf 91.75 & 57.75 & 26.05 & \bf 44.01 & 43.41 & 86.48 & \bf 97.29  & 91.89 \\
Random forest & 77 & \bf 93.5 & 69.75 & 29.94 & \bf 54.19 & 51.80& 97.29 & \bf  100 & 91.89 \\
\bottomrule
\end{tabular}
}
\label{acc_rep}
\end{table}

\subsection{Evaluating embedded-document relevance}\label{section:eval_embed_yelp}
\subsubsection{On binary classification task (Yelp review and CDD reports)}

To measure the quality of attribution scores, in this experiment, the columns in the embedded documents (referred as features) are gradually removed. While removing features with the largest (Table~\ref{rm_features_dv}) or smallest absolute (Table~\ref{rm_features_dv_irr}) attribution scores, the model accuracy was recorded to assess whether the truly relevant features have been identified. In Table~\ref{rm_features_dv}, LRP resulted in larger decrease in model accuracy by removing the most relevant features. In Table~\ref{rm_features_dv_irr}, compared to random feature removal, LRP and SA were both able to preserve the accuracy by removing the least relevant features.


\begin{table}[!ht]
\centering
\caption{Accuracy score (\%) on binary classification task (with TextCNN) after removing relevant features of documents}
\resizebox{\textwidth}{!}{
\begin{tabular}{@{} l ccc ccc ccc ccc @{}}
\toprule
Nr-removal & \multicolumn{6}{c}{Yelp reviews} &  \multicolumn{6}{c}{CDD reports} \\ 
 & \multicolumn{3}{c}{Positive (class ``1'')} &  \multicolumn{3}{c}{Negative (class ``0'')} & \multicolumn{3}{c}{High risk (class ``1'')} &  \multicolumn{3}{c}{Low risk (class ``0'')} \\ 
 & Rand & SA  & LRP & Rand & SA  & LRP & Rand & SA  & LRP & Rand & SA  & LRP \\ 
\midrule
 50 & 99.5 & 98 & \bf 44.5 &  99.05 & 98.40 &  99.05 & 100 & 100 & \bf 98 & 100 & 95.18 & \bf 91.56 \\
 100 & 96.3 & 97.9 & \bf 6.7 & 96.6 & 98.85 & 97.7 & 98.79 & 100 & \bf 96 & 100 & 95.18 & \bf 86.75 \\
 150 & 92.7 & 96.4 & \bf 2.1 & 92.35 &  98.05 & 95.65 & 98.79 & 100 & \bf 89 & 100 & 95.18 & \bf 6.02 \\
\bottomrule
\end{tabular}
}
\label{rm_features_dv}
\end{table}

\begin{table}[!ht]
\centering
\caption{Accuracy score on binary classification task (with TextCNN) after removing irrelevant features of documents}
\resizebox{\textwidth}{!}{
\begin{tabular}{@{} l ccc ccc ccc ccc @{}}
\toprule
Nr-removal & \multicolumn{6}{c}{Yelp reviews} &  \multicolumn{6}{c}{CDD reports} \\ 
 & \multicolumn{3}{c}{Positive (class ``1'')} &  \multicolumn{3}{c}{Negative (class ``0'')} & \multicolumn{3}{c}{High risk (class ``1'')} &  \multicolumn{3}{c}{Low risk (class ``0'')} \\ 
 & Rand & SA  & LRP & Rand & SA  & LRP & Rand & SA  & LRP & Rand & SA  & LRP \\ 
\midrule
50 & 98.4 & 99.4 & \bf 99.6 & 95.2 & \bf 99.9 & \bf 99.9 & 97.59 & \bf 100 & \bf 100 & 98.79 & 96.39 & \bf 100 \\
100 & 98.8 & \bf 99.3 & 98.8 & 96.1 & 99.5 & \bf 99.8 & \bf  100 & \bf 100 & \bf 100 & 96.39 & 92.77 & \bf 100 \\
150 & 98.5 & 98.6 & 98.7 & 94 & 96.6 & \bf 99.7 & 95.18 & \bf 100 & \bf 100 & 93.97 & 96.38 & \bf 100 \\
\bottomrule
\end{tabular}
}
\label{rm_features_dv_irr}
\end{table}

\subsubsection{On multi-class classification task (US customer financial complaints)}\label{section:eval_embed_us}

\begin{figure}[!ht]
\centering
\begin{subfigure}{0.45\textwidth}
\includegraphics[width=\linewidth]{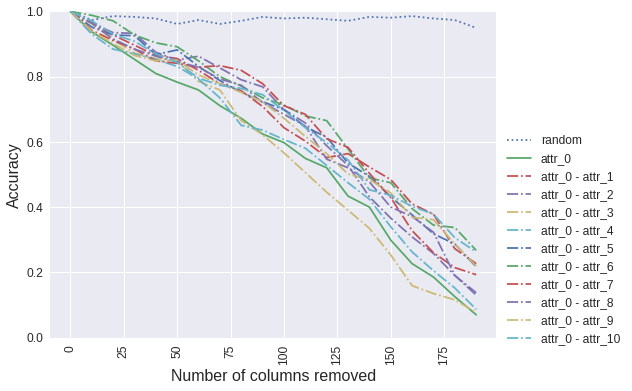}
\end{subfigure}
\begin{subfigure}{0.45\textwidth}
\includegraphics[width=\linewidth]{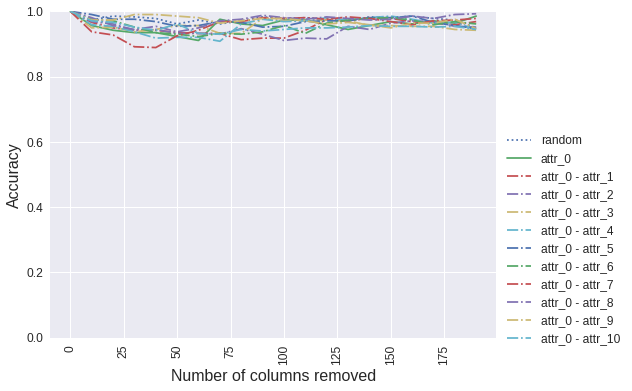}
\end{subfigure}
\caption{Accuracy when embedding columns with the largest LRP (left) and saliency (right) attribution scores/attribution differences are removed on US consumer financial complaints dataset}
\label{fig:embedding_acc}
\end{figure}

In this experiment, 415 documents that were correctly classified as class "0" (bank account or service) were investigated. Based on both LRP and saliency attribution scores, as well as the attribution differences between actual class and other classes, we gradually removed embedding columns with the largest relevance. Figure~\ref{fig:embedding_acc} shows the changes in model accuracy. A significant decline in model accuracy can be observed for the LRP attributions. When using the saliency approach, the accuracy change is similar to random feature removal. 

\begin{table}[!ht]
\vspace{-1em}
\centering
\caption{Number of mis-classified documents for chosen actual class ($0$) and two other classes ($2$ and $3$)}
\begin{tabular}{c c cc cc cc} 
\hline
Predictions & Nr. columns removed & \multicolumn{2}{c}{attr\_0} & \multicolumn{2}{c}{attr\_0 - attr\_2} & \multicolumn{2}{c}{attr\_0 - attr\_3} \\
 &  & LRP & SA & LRP & SA & LRP & SA \\ 
\hline
0 & 50 & 325 & 383 & 353 & 387 & 352 & 397   \\
& 100 & 248 & 404 & 267 & 378 & 235 & 403    \\
& 150  & 124 & 402 & 152  & 400 & 105 & 397   \\ 
\hline
2 & 50  & 21  & 9 & \textbf{27} & 5 & 18 & 5     \\
& 100  & 33  & 5  & \textbf{41} & 12 & 40  & 5    \\
& 150 & 36  & 6  & \textbf{86} & 9  & 41 & 4      \\ 
\hline
3 & 50  & 4   & 3 & 4 & 2 & \textbf{7}   & 0         \\
& 100  & 9   & 1  & 6 & 2 & \textbf{29}  & 0       \\
& 150  & 37  & 2  & 27 & 0  & \textbf{100} & 0      \\
\hline
\end{tabular}
\label{table:prediction_count_1}
\end{table}

The accuracy decrease was also observed by using perturbation based on the LRP attribution differences. To investigate how the model prediction is altered based on attribution differences, the number of mis-classifications in each class were recorded while the embedding columns are gradually removed, as shown in Table~\ref{table:prediction_count_1}. Documents that are correctly classified as class "0" (\textit{bank account or service}) is used as a baseline (``attr$_0$''). To investigate the role of attribution differences, we choose an example of attribution differences between true class ``0'' and class ``2'' (\textit{credit card}) (``attr$_0-$attr$_2$''), and between actual class ``0'' and class``3'' (\textit{credit reporting}) (``attr$_0-$attr$_3$''). 

When attributions towards the true class were used, the number of documents correctly classified was smaller with the LRP approach, which is consistent with the results presented in  Figure~\ref{fig:embedding_acc}. What is worth noticing is that when using the LRP attribution differences, the prediction is guided towards favoring a certain class. When applying attribution differences between true class and class ``2'', for instance, the number of documents mis-classified as ``2'' is significantly larger than using other feature removal metrics. We make the same observation with the attribution differences between true class and class ``3''. This shows that we could also use the attribution differences removal method, in addition to removing largest and smallest relevance score, to evaluate the quality of attribution methods.    

\subsection{Evaluating embedded-$n$-grams relevance}
\subsubsection{On binary classification task (Yelp review and CDD reports)}\label{section:eval_ngram_yelp}

\begin{table}[!ht]
\centering
\caption{Accuracy score on the binary classification task after removing relevant and irrelevant $n$-grams features}
\resizebox{\textwidth}{!}{
\begin{tabular}{@{} c ccc ccc ccc ccc @{}}
\toprule
Nr-removal & \multicolumn{6}{c}{Remove relevant features} & \multicolumn{6}{c}{Remove irrelevant features} \\
& \multicolumn{3}{c}{Yelp reviews} &  \multicolumn{3}{c}{CDD reports} & \multicolumn{3}{c}{Yelp reviews} &  \multicolumn{3}{c}{CDD reports} \\ 
  & Rand & SA  & LRP & Rand & SA  & LRP & Rand & SA  & LRP & Rand & SA  & LRP \\ 
\midrule
1 & 99.9 & 98.6 & \bf 98.3 & 99.45 & 100 & 100 & 99.7 & 99.8 & \bf 99.9 & 100 & 100 & 100  \\
3 & 99.7 & 92.7 & \bf 91.2 & 99.45 & 98.90 & 97.81 & 99.1 & \bf  99.8 & \bf  99.8 & 100 & 100 & 100  \\
5 & 99.2 & 81.1 & \bf 78.8 & 100 & 97.81 & 96.17 & 99.1 & \bf  99.7 & 99.4 & 100 & 98.9 & 100  \\
7 & 99.3 & 63 & \bf 59.7 & 98.9 & 93.98 & 89.07  & 99.1 & 98.75 & \bf  99.45 & 100 & 100 & 100  \\
\bottomrule
\end{tabular}
}
\label{rm_features_ng}
\end{table}

Table~\ref{rm_features_ng} invites us to make similar observations as in Section~\ref{section:eval_embed_yelp}, but by using the convolutional filter feature removal method. By removing relevant features, larger impact on the model accuracy was resulted in LRP-based approach. Likewise, by removing irrelevant features, both LRP and SA were able to preserve model accuracy compared to the random feature removal. 

\subsubsection{On multi-class classification task (US consumer financial complaints)}\label{section:eval_ngram_us}

Similar to the procedure described in Section~\ref{section:eval_embed_us}, only the documents that were correctly classified were investigated. Instead of removing relevant embedding columns, convolutional filters were regarded as the feature to be assessed. In both the LRP and saliency approaches, the model accuracy decreased drastically as $n$-gram influences on certain positions were removed from the model. To investigate whether the predictions were guided towards a certain class, the number of mis-classifications for each class is also recorded in Table~\ref{table:prediction_count_2}. While the feature removal based on attributions of the true class was able to alter the predictions towards class ``2'' and ``3'', the mis-classification numbers were significantly higher when using both the LRP and saliency attribution differences. The predictions were indeed guided towards desired classes.

\begin{figure}[!ht]
\centering
\begin{subfigure}{0.45\textwidth}
\includegraphics[width=\linewidth]{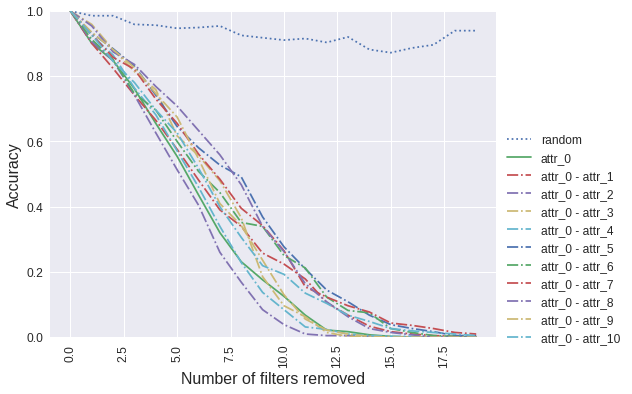}
\end{subfigure}
\begin{subfigure}{0.45\textwidth}
\includegraphics[width=\linewidth]{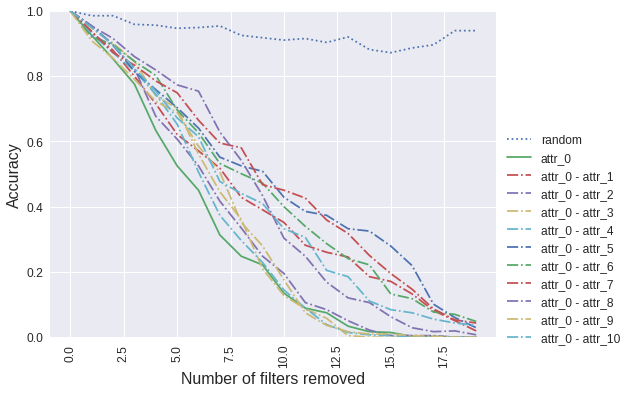}
\end{subfigure}
\caption{Accuracy when convolutional filters with the largest LRP(left) and saliency(right) attribution scores/attribution differences are removed on US consumer financial complaints dataset}
\label{fig:conv_acc}
\end{figure}

\begin{table}[!ht]
\vspace{-1em}
\centering
\caption{Number of mis-classified documents for the actual class ($0$) and two other classes ($2$ and $3$)}
\begin{tabular}{c c cc cc cc} 
\hline
Predictions & Nr. filters removed & \multicolumn{2}{c}{attr\_0} & \multicolumn{2}{c}{attr\_0 - attr\_2} & \multicolumn{2}{c}{attr\_0 - attr\_3} \\
 &  & LRP & SA & LRP & SA & LRP & SA \\ 
\hline
0 & 5 & 239 & 218 & 213 & 252 & 256 & 291 \\
 & 10 & 93 & 146 & 16 & 81 & 40 & 53 \\
 & 15 & 7 & 71 & 0 & 2 & 0 & 0 \\ 
\hline
2 & 5 & 46 & 83 & \bf 121 &\bf 102 & 19 & 13 \\
 & 10 & 69 & 104 & \bf 317 &\bf 290 & 24 & 19 \\
 & 15 & 42 & 87 & \bf 367 & \bf 389 & 12 & 10 \\ 
\hline
3 & 5 & 14 & 8 & 8 & 6 &\bf 31 &\bf 22 \\
 & 10 & 44 & 16 & 8 & 5 &\bf 145 &\bf 195 \\
 & 15 & 83 & 23 & 7 & 6 &\bf 287 &\bf 319 \\
\hline
\end{tabular}
\label{table:prediction_count_2}
\end{table}

\section{Conclusion}
\label{sec:conclude}

In this paper, we presented an experimental study on feature-based perturbations for evaluating attribution-based explanations on CNN model for text classification (TextCNN). Instead of utilizing ``word-deleting'' evaluation, we investigated the attribution-based explanations on different layers of TextCNN. Our experimental analysis was performed on two public data sets (Yelp reviews and US customer complaints) and extracted customer reports from CDD cases of a financial institution, by using three different aspects of attribution scores: the embedded word level, the embedded document level, and the embedded $n$-gram level. Our proposed evaluation was able to assess the quality of attribution scores with a measurable metric, while showing the differences in different explanation approaches. The results of our experimental study suggest that LRP is better at finding features that are relevant to the prediction. By investigating the attribution differences, we were also able to analyze whether the model's prediction is guided to a certain outcome. We provided a visualization tool to offer deeper insights into the model's predictions by visualizing the LRP attributions as well as the attribution differences between different classes on individual words and $n$-grams.

\bibliography{references}

\end{document}